\RequirePackage{fix-cm}
\documentclass[smallextended]{svjour3}       % onecolumn (second format)
\smartqed  % flush right qed marks, e.g. at end of proof
\usepackage{graphicx}
\usepackage{color}
\usepackage{subfig}
\usepackage{dashbox}
\usepackage{indentfirst}
\usepackage{amsmath}
\usepackage[colorlinks,bookmarksopen,bookmarksnumbered,citecolor=black, urlcolor=black]{hyperref}

\usepackage[most]{tcolorbox}

\definecolor{custom-gray}{cmyk}{0, 0, 0, 0.7, 1.00}
\newtcbtheorem[no counter]{Summary}{\hskip-1.2em}{enhanced,drop shadow={black!50!white},
  coltitle=white,
  top=0.15in,
  attach boxed title to top left=
  {xshift=1.5em,yshift=-\tcboxedtitleheight/2},
  boxed title style={size=small,colback=custom-gray}
}{summary}

\begin{document}

\title{
Can We Recycle Our Old Models? An Empirical Evaluation of Model Selection Mechanisms for AIOps Solutions
% model selection  historical 
%about the article that should go on the front page should be
%placed here. General acknowledgments should be placed at the end of the article.}
}

\author{Yingzhe~Lyu         \and
        Hao~Li \and
        Heng~Li \and
        Ahmed~E.~Hassan
}

%\author{Yingzhe~Lyu}
%\email{ylyu@cs.queensu.ca}
%\affiliation{%
%  \institution{Queen's University}
%  \department{Software Analysis and Intelligence Lab (SAIL)}
%  \city{Kingston}
%  \state{ON}
%  \country{Canada}
%}

%\author{Heng~Li}
%\email{heng.li@polymtl.ca}
%\affiliation{%
%  \institution{Polytechnique Montreal}
%  \department{Department of Computer and Software Engineering}
%  \city{Montreal}
%  \state{QC}
%  \country{Canada}
%}

%\author{Ahmed~E.~Hassan}
%\email{ahmed@cs.queensu.ca}
%\affiliation{%
%  \institution{Queen's University}
%  \department{Software Analysis and Intelligence Lab (SAIL)}
%  \city{Kingston}
%  \state{ON}
%  \country{Canada}
%}

\institute{
    Yingzhe~Lyu, Hao~Li, Ahmed~E.~Hassan  \at
    Software Analysis and Intelligence Lab (SAIL) \\
    Queen's University, Kingston, Ontario, Canada \\
    \email{\{ylyu, ahmed\}@cs.queensu.ca, hao.li@queensu.ca}  \\
    \and
    Heng~Li \at
    Department of Computer and Software Engineering \\
    Polytechnique Montreal, Montreal, Quebec, Canada \\
    \email{heng.li@polymtl.ca}
}
% \institute{F. Author \at
%               first address \\
%               Tel.: +123-45-678910\\
%               Fax: +123-45-678910\\
%               \email{fauthor@example.com}           %  \\
% %             \emph{Present address:} of F. Author  %  if needed
%            \and
%            S. Author \at
%               second address
% }

\date{Received: date / Accepted: date}
% The correct dates will be entered by the editor

\maketitle

\begin{abstract}
AIOps (Artificial Intelligence for IT Operations) solutions leverage the tremendous amount of data produced during the operation of large-scale systems and machine learning models to assist software practitioners in their system operations.
%As operation data produced in the field are constantly evolving due to factors such as the changing operational environment and user base, the models used in AIOps solutions need to be constantly maintained after deployment.
Existing AIOps solutions usually maintain AIOps models against concept drift through periodical retraining, despite leaving a pile of discarded historical models that may perform well on specific future data. 
Other prior works propose dynamically selecting models for prediction tasks from a set of candidate models to optimize the model performance. % as well as mitigate the issue of concept drift.
However, there is no prior work in the AIOps area that assesses the use of model selection mechanisms on historical models to improve model performance or robustness.
To fill the gap, we evaluate several model selection mechanisms %on historical models in AIOps solutions 
by assessing their capabilities in selecting the optimal AIOps models that were built in the past to make predictions for the target data. % in history for prediction, and assess these model selection mechanisms' prediction performance and the overall ranking performance.
We performed a case study on three large-scale public operation datasets: two trace datasets from the cloud computing platforms of Google and Alibaba, and one disk stats dataset from the BackBlaze cloud storage data center.
We observe that the model selection mechnisms utilizing temporal adjacency tend to have a better performance and can prevail the periodical retraining approach.
Our findings also highlight a performance gap between existing model selection mechnisms and the theoretical upper bound which may motivate future researchers and practitioners in investigating more efficient and effective model selection mechanisms that fit in the context of AIOps.

\end{abstract}

\section{Introduction}
\label{sec:intro}
% introduction to AIOps
Modern large-scale software systems can generate tremendous amounts of data during their daily operations. 
As the volume of operation data grows, it is getting increasingly challenging for practitioners to manually analyze and utilize such data. 
AIOps (\underline{A}rtificial \underline{I}ntelligence for IT \underline{Op}eration\underline{s}), which helps practitioners automatically leverage such rich information on the system's runtime behavior and health condition~\cite{gartner2018aiops}, has gained increasing popularity among practitioners and researchers.
Today, AIOps solutions support various goals in software and system operations, such as machine failure predictions~\cite{lin2018predicting,li2020predicting}, job failure predictions~\cite{elsayed2017learning,rosa2015catching}, disk failure predictions~\cite{xu2018improving,botezatu2016predicting,elsayed2017learning}, performance degradation detection~\cite{chaudhary2023monitoring,liao2025early}, and service outage predictions~\cite{chen2019outage}.
Many proposed AIOps solutions have already demonstrated promising benefits in practice~\cite{li2020predicting,li2018adopting,bansal2020decaf,weng2022mlaas,bendimerad2023premise}.
For example, Lin et al.~\cite{lin2018predicting} successfully applied their technique on one of Microsoft's large-scale cloud service systems to predict potential node failures based on historical data. 
Botezaku et al.~\cite{botezatu2016predicting} leverage monitoring information to build a machine learning pipeline for predicting hard drive failures on large-scale cloud computing platforms.

% Concept drift challenges facing AIOps maintenance
Despite advances in ML models and their applications in AIOps, many challenges are still associated with the maintenance and data evolution of AIOps solutions following their deployment in the field.
%\heng{wild or field? same for the use of wild below}.
Concept drift~\cite{wang2010mining,wang2003mining,tsymbal2004problem,nishida2007detect}, which refers to the change in data distribution and in the relationship between the variables, can lead to the obsolescence of models trained on stale data and negatively impact the performance in the future time~\cite{li2020predicting}.
Our prior works~\cite{lyu2021empirical,lyu2024update} find that the operation datasets are subject to a considerable scale of concept drift, negatively affecting the model performance and stability in the field.
In order to mitigate the issue of concept drift, practitioners should conduct constant maintenance and update to sustain the model performance and stability once deployed in the field~\cite{lyu2021empirical,lyu2024update,dang2019aiops}.
However, existing AIOps studies either train a static model regardless of the potential threat from concept drift~\cite{elsayed2017learning,rosa2015catching,botezatu2016predicting,mahdisoltani2017proactive,chen2019outage,xue2018spatial} or periodically retrain the model to maintain performance and stability~\cite{li2020predicting,dang2019aiops,lin2018predicting}.
Stationary models may suffer from performance degradation and impact user experience in the field, while periodically retraining the model could be very expensive (e.g., resources and human efforts involved in updating, verifying, and integrating the model)~\cite{li2020predicting}.

% bringing in the model selection mechanisms
As reported in previous work, ensuring the quality of software systems is still an open challenge for the research community~\cite{openja2025empirical}, and we set out to find more efficient tools to maintain AIOps solutions.
Previous works~\cite{diffenderfer2021winning,zhang2025dynamic} suggest that model decision-making mechanisms can benefit model performance and robustness when faced with concept drift, as they leverage the diversity of models to provide robust predictions in dynamic and changing conditions.
%The employment of such model selection mechanisms has been observed in various service applications.
%For example, AWS fraud detection~\cite{aws} trains an auxiliary unsupervised anomaly detection model in addition to the main supervised model to augment the prediction results.
%IBM Watson natural language understanding~\cite{ibm} also uses an ensemble learning framework to include predictions from multiple models.
As AIOps models deployed in the field require constant retraining to maintain their performance and steer away from obsolescence~\cite{li2018adopting,lin2018predicting,li2020predicting,xu2018improving}, a large number of historical models would accumulate during this process.
Some useful information will likely still be buried in these historical models and can be salvaged and recycled to improve prediction performance and overall performance stability. %\heng{do we evaluate stability/robustness? if not we don't need to motivate it here}.
However, there is no prior work and little empirical evidence in the AIOps area that assesses the use of model selection mechanisms in improving model performance and robustness. %\heng{same here}.
To fill the gap, we propose to examine the application of model selection mechanisms on the historical models for improving model performance and robustness %\heng{make sure robustness is evaluated if mentioned here} 
while introducing minimal cost, namely \emph{model selection mechanisms on historical models}.
We conduct a case study to evaluate various forms of selection mechanisms on historical models in AIOps solutions that select the best models in history to make predictions, and assess these model selection mechanisms' performance and robustness on operation datasets.
This study aims to address the following research questions:%\heng{minor adjustment to RQs}

\begin{itemize}
    %\item How do the historical models perform on the new data
    \item RQ1: How well can the model selection mechanisms achieve optimal performance for AIOps solutions?%~\yingzhe{Please check our new results, I feel like since we have enough description of oracle in experiment design as well as comparison in the experiment results, having a RQ0 on motivating the oracle would be repetitive.}\heng{how about the heatmap we discussed last time? I think that could be an interesting contribution of the work: it tells what the distances are between time windows and the optimal historical models.}~\yingzhe{We now have 4 models on 3 datasets which means there would be 12 heatmaps in the results (also the Alibaba one would be a 4x4 upper-triangle matrix with no cell shows better performance than retraining), would be hard to present.}
    \item RQ2: How well can the model selection mechanisms achieve optimal model ranking for AIOps solutions?
    \item RQ3: How stable are the model ranking results achieved by the model selection mechanisms? % in AIOps solutions?\heng{is this done?}
\end{itemize}

Our main contribution includes proposing the application of model selection mechanisms to select historical AIOps models. % and its historical models.
We establish the theoretical performance upper bound for model selection mechanisms on historical models %when predicting testing samples with the top-ranked model 
through a hypothetical oracle.
We are also the first to conduct an empirical study that evaluates historical model selection mechanisms on AIOps solutions.
In addition, We share a replication package which includes our code for conducting case study on the studied operation datasets and analyzing the experiment results,\footnote{\url{https://github.com/EmpyreanKnight/suppmaterial-25-yingzhe-AIOpsSelection}} so that others in the research community can replicate or extend our work.

The paper is organized as follows.
Section~\ref{sec:background} presents an overview of prior works in AIOps solution and model selection mechanisms.
Section~\ref{sec:experimental_design} details our case study design and methodology for evaluating model selection mechanisms on historical models in the area of AIOps.
Section~\ref{sec:results} delivers our experiment results and analysis.
Section~\ref{sec:threats} discusses the limitations of our work and possible threats to validity.
Finally, Section~\ref{sec:conclusion} concludes our work.

\section{Related Work}
\label{sec:background}
\subsection{Prior research on AIOps solutions}

Although huge efforts have been devoted to large-scale software systems like cloud computing to ensure the quality of services, various types of operational incidents (e.g., job termination, hard drive failure, and performance anomalies) are still unavoidable.
To ensure the reliability of uninterrupted services, operational incidents must be identified, resolved, and managed in a timely manner, as failing to do so could interrupt service availability and incur massive financial damage~\cite{zhang2024survey}.
Prior works have proposed various AIOps solutions for addressing different problems in the operation of large-scale software and systems, including incident prediction~\cite{lin2018predicting,li2020predicting,elsayed2017learning,rosa2015catching,botezatu2016predicting,mahdisoltani2017proactive,xu2018improving,chen2019outage}, anomaly detection~\cite{he2018identifying,lim2014identifying}, ticket management~\cite{xue2018spatial,xue2016managing}, issue diagnosis~\cite{luo2014correlating}, and self healing~\cite{ding2012healing,ding2014mining,lou2013software,lou2017experience}.
These AIOps solutions can be categorized into two phases that contribute to incident management: 
1) \emph{incident perception}, which predicts whether certain types of incidents would occur by learning from the historical operation data; 
and 2) \emph{incident mitigation}, which remediates the damage from incidents (e.g., automated problem diagnosis) or provides suggestions to domain experts (e.g., incident triage) after the incident occurrences.

\subsubsection{Incident perception}
Incident perception is a vital step towards early detection of potential incidents and proactive prevention of system failures.
Prior works have focused on the identification of various types of incidents by analyzing monitoring data.
These approaches can be categorized into two types: failure prediction~\cite{elsayed2017learning,rosa2015catching,botezatu2016predicting,mahdisoltani2017proactive,li2014failure,li2017failure,xu2018improving,zhao2020incident,chen2019outage,lin2018predicting,li2020predicting,yang2023diffusion,alharthi2023time} and anomaly detection~\cite{zhang2022deep,zeng2023traceark,wu2023effectiveness,yang2021semi,almodovar2024logfit}.

\textbf{Failure prediction}.
Failure prediction involves forecasting potential system failures before their occurrence by analyzing historical data and identifying patterns that precede failures.
For example, El-Sayed et al.~\cite{elsayed2017learning} and Rosa et al.~\cite{rosa2015catching} predict job failures from trace data collected from the Google cloud computing platform. 
Botezaku et al.~\cite{botezatu2016predicting}, Mahdisoltani et al.~\cite{mahdisoltani2017proactive}, Li et al.~\cite{li2014failure,li2017failure}, and Xu et al.~\cite{xu2018improving} leverage SMART-based monitoring data to build a machine learning pipeline for predicting hard drive failures in large-scale cloud computing platforms. 
Zhao et al.~\cite{zhao2020incident} propose a deep-learning-based approach, \emph{eWarn}, which leverages textual (e.g., keywords in incident tickets) and statistical features (e.g., alert count) to predict incident occurrences. 
Similarly, Chen et al.~\cite{chen2019outage} collect and analyze the alert data and its dependencies to predict outages in the whole cloud system.
Lin et al.~\cite{lin2018predicting} and Li et al.~\cite{li2020predicting} predict node failures in large-scale cloud computing platforms by building machine learning models from temporal (e.g., CPU utilization metrics), spatial (e.g., location of a node), and config data (e.g., build information).  
Yang et al.~\cite{yang2023diffusion} propose a novel diffusion model that enhances data quality through data imputation to improve the performance of the downstream failure prediction task in the cloud scenario.
Alharthi et al.~\cite{alharthi2023time} propose a two-stack transformer-decoder architecture to predict failures as well as the lead times in HPC systems.

\textbf{Anomaly detection}.
Anomaly detection aims to detect abnormal system behaviors or patterns that indicate potential issues or failures to help developers and operators uncover system issues and solve anomalies.
For instance, Zhang et al.~\cite{zhang2022deep} propose a deep-learning-based microservice anomaly detection approach that uses a unified graph representation to describe the complex structure of a trace together with log events embedded in the structure. 
Zeng et al.~\cite{zeng2023traceark} propose an actionable performance anomaly alerting approach based on trace data for online service systems. 
Wu et al.~\cite{wu2023effectiveness} conduct a comprehensive evaluation of seven supervised anomaly detection models with six log representations on four public datasets.
Yang et al.~\cite{yang2021semi} propose a semi-supervised approach for detecting log-based anomalies that can stay immune to unstable log data via semantic embedding.
Almodovar et al.~\cite{almodovar2024logfit} propose a fine-tuned language model that is robust to log content change, \emph{LogFiT}, for anomaly detection in the HuggingFace ecosystem. 

\subsubsection{Incident mitigation}
Incidents in software systems need to be mitigated in a timely manner. 
Prior works have focused on triaging~\cite{chen2019continuous,chen2020incidental,bansal2020decaf,he2018identifying}, diagnosing~\cite{zhang2005ensembles,luo2014correlating,banerjee2020challenges,jehangiri2014diagnosing,chen2020identifying,wang2023root,misiakos2023dag}, and managing issues~\cite{lou2013software,lou2017experience,xue2016managing,xue2018spatial,jiang2020mitigate,lin2016idice,lim2014identifying}, which benefit the mitigation process. 

\textbf{Triaging}.
Chen et al.~\cite{chen2019continuous} propose a deep-learning-based technique to improve the current incident triage process (e.g., distributing the new incident to the responsible team). 
Chen et al.~\cite{chen2020incidental} perform an empirical study on characterizing incidents in online systems and propose \emph{DeepIP}, a technique to detect incidental incidents (i.e., incidents that are less severe and last for a short period of time), which can reduce the incident triage efforts. 
Bansal et al.~\cite{bansal2020decaf} propose \emph{DeCaf}, a Random Forest-based framework to correlate telemetry data with performance regressions. In addition, the detected performance regressions are automatically triaged to the on-site engineering team. 

\textbf{Diagnosing}.
Zhang et al.~\cite{zhang2005ensembles} propose an ensemble of models to automatically diagnose performance problems. 
Chen et al.~\cite{chen2020identifying} propose \emph{LiDAR}, a deep-learning-based approach to linking similar incidents based on historical information. 
Luo et al.~\cite{luo2014correlating} mine time-series data and event data to discover correlations between them, which could improve the incident diagnosis process. 
Banerjee et al.~\cite{banerjee2020challenges} discuss challenges in performance diagnosis in a hybrid-cloud enterprise software environment. 
Jehangiri et al.~\cite{jehangiri2014diagnosing} present techniques to diagnose performance anomalies using time-series datasets.

\textbf{Managing}. 
Jiang et al.~\cite{jiang2020mitigate} analyze the similarity between incident descriptions and their corresponding troubleshooting guide to facilitate incident management. 
Lou et al.~\cite{lou2013software,lou2017experience} develop a software analytic-based system to resolve the scalability, reliability, and maintainability of data-driven incident management systems. 
Lim et al.~\cite{lim2014identifying} leverage performance metrics to cluster performance issues into recurrent and unknown ones.
Xue et al.~\cite{xue2016managing,xue2018spatial} proactively reduce performance tickets by predicting usage series in cloud data centers. 
Lin et al.~\cite{lin2016idice} propose a data mining-based technique to detect emerging issues (a sudden burst of new issues) by analyzing historical issues.

\subsection{Prior work on model selection techniques}
% introduce the model selection techniques and why they are needed.
%\yingzhe{The structure is now general model selection $\rightarrow$ multiple candidate historical models available $\rightarrow$ selection can be useful here too. Maybe too short? Add more selection techniques later. }\heng{Overall, the flow looks good}
Various machine learning models have been proposed for different inference and prediction tasks.
However, no existing model can accommodate every type of data and goal.
Model selection mechanisms mitigate the challenge by considering a set of candidate models and selecting the most appropriate ones for the prediction task~\cite{ding2018selection}.
For example, Liu et al.~\cite{liu2022consistent} propose Consistent Relative Confidence (CRC), a label-free model selection method using only unlabeled testing data based on the confidence of candidate models.
The authors find that there is a strong positive relationship between consistent relative confidence and correctness among a group of rational volunteers: if there is one candidate who acts more confidently than the others in a consistent way, then its decision tends to be more accurate than the others.
Hu et al.~\cite{hu2024laf} propose a labeling-free model selection approach against performance degradation on out-of-distribution data.
The main idea is to statistically learn a Bayesian model with the expectation-maximization (EM) algorithm to infer the models' specialty only based on predicted pseudo-labels.

% Further discuss the model selection and it's usability in concept drift/AIOps scenario
In the context of continual or lifelong learning~\cite{parisi2019continual}, a great number of candidate models can be available from previous updates, and choosing the right one with the best generalization property for the current task remains a tough challenge.
Given a set of candidate models, there is no prior evidence of which model will be capable of solving the targeted task the most effectively~\cite{hu2024laf}.
Prior work reports that model selection mechanisms can benefit model performance and robustness when facing concept drift as they leverage the diversity of models to provide robust predictions in dynamic and changing conditions~\cite{diffenderfer2021winning,zhang2025dynamic}.
When predicting the test samples, model selection mechanisms can estimate the model performance and rank the models to pick models that most likely to perform best. 
Research in the AIOps area also suffers from the challenge of concept drift and usually relies on periodical retraining to maintain model performance~\cite{li2020predicting,dang2019aiops,lin2018predicting}.
%\heng{Also discuss prior work on AIOps model maintenance such as your TOSEM paper}\yingzhe{our maintenance paper is more focused on retraining more smartly rather than improving performance}
The historical models being discarded may contain valuable information and perform well on specific future periods~\cite{poenaru2024your}.
However, there is no prior work that systematically examines the performance of model selection mechanisms on historical models in the context of AIOps.

\section{Experiment Design}
\label{sec:experimental_design}

\subsection{Case study setup}

In order to evaluate different mechanisms of model selection using historical models in the context of AIOps, we perform a case study on three large-scale operation datasets: the Google cluster trace dataset~\cite{google2011cluster}, the Backblaze disk stats dataset~\cite{backblaze2020drive}, and the Alibaba GPU cluster trace dataset~\cite{alibaba2020cluster}.
We choose to carry out a case study on these three datasets for the following reasons: 1) they are publicly available; 2) they are large-scale datasets and cover relatively long operation periods (i.e., months to years), which enables us to examine model selection mechanisms over the evolution of the data.
In addition, prior works have widely studied the first two datasets in particular for predicting job failures on the Google dataset~\cite{elsayed2017learning,rosa2015predicting} and predicting disk failures on the Backblaze dataset~\cite{botezatu2016predicting,xu2018improving,mahdisoltani2017proactive}.
In this work, we focus on predicting job outcomes (i.e., failure or not) on the Google and Alibaba cluster trace datasets and predicting disk failures on the Backblaze disk stats dataset as done in prior works~\cite{lyu2021empirical,lyu2022towards,lyu2024update}.

\subsubsection{Google cluster trace dataset}

\begin{figure}[!htbp]
    \centering
    \subfloat[Google data schema.]{
        \dbox{\includegraphics[width=0.2812\textwidth]{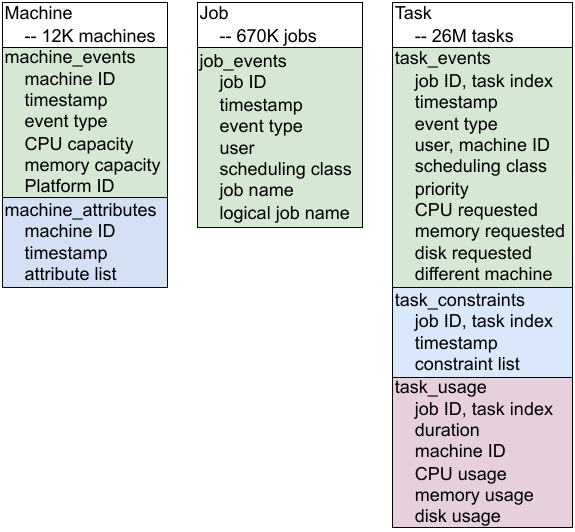}}
        \label{fig:google_schema}
    }
    \subfloat[Backblaze data schema.]{
        \dbox{\includegraphics[width=0.2565\textwidth]{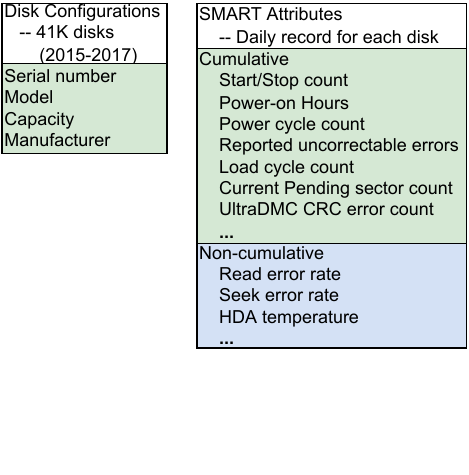}
        }
    \label{fig:disk_schema}
    }
    \subfloat[Alibaba data schema.]{
        \dbox{\includegraphics[width=0.3173\textwidth]{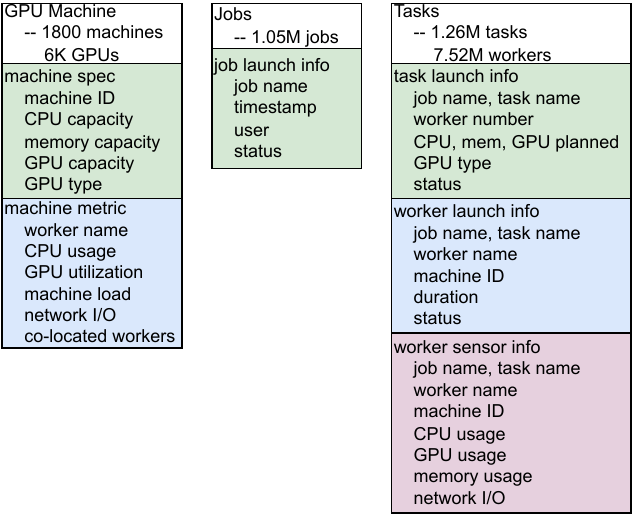}
        }
    \label{fig:alibaba_schema}
    }
    \caption{Data schema for our studied datasets. 
    Each colored box represents a data table: a line of the table name followed by lines describing the data fields.
    For the Google and Alibaba datasets, each table (e.g., machine\_events) is one or multiple CSV files containing the fields described in the box. 
    For the Backblaze dataset, the tables represent the logical view, while the physical data is stored as daily snapshots of each disk's attributes.
    }
    \label{fig:schema}
\end{figure}

The cluster data released by Google in 2011 contains the trace data of a production cluster with about 12K machines in 29 days for 670K jobs and 26M tasks~\cite{chen2014failure}. 
The data features workload arrives at a cell (i.e., a set of machines that share a common cluster-management system) in the form of jobs. Each job comprises one or more tasks, and each task is scheduled on a single machine.
Figure~\ref{fig:google_schema} shows the dataset schema and information provided in the Google cluster trace dataset.

Following prior works~\cite{elsayed2017learning,rosa2015predicting}, our goal on the Google cluster data is to predict whether a job will fail or not (i.e., terminated for any reason before successfully completed) using the information at job submission and the monitoring data in the first five minutes of the job execution. 
In the Google cluster trace dataset, each job has several events, and each is associated with a transit (e.g., submit, schedule, evict, fail, kill, finish, lost, update) among the states (e.g., unsubmitted, pending, running, dead) in the job's lifecycle. 
We consider a job fails if its final state is ``fail'', same as in prior works~\cite{elsayed2017learning,rosa2015predicting}.

Similarly to El-Sayed et al.~\cite{elsayed2017learning}, we predict job failures using the configuration and temporal features.
Configuration features are values determined upon job submission, such as the requested CPU, memory, and disk space.
In contrast, temporal features are values that change during a job's execution, such as the mean and standard deviation of CPU, memory, and disk space usage by a job over the first five minutes since job submission.
%The detailed list of the features for job failure prediction is described in Table~\ref{tab:features_google}, where the first 9 are configuration features while the latter 6 are temporal features. 

We remove the jobs that are not completed or whose records are lost during execution, as the final states of these jobs are missing.
We further remove the jobs that start on the last day (i.e., the 29th day), as these jobs are more likely to remain incomplete before the data cutoff time and cause data collection bias (completed jobs typically last longer than the failed ones). In fact, we observed a much higher job failure rate from the jobs starting on the last day.   
In addition, we removed jobs that finished in less than five minutes since their submission as they have not generated sufficient metrics for prediction. 
We also observe that a large proportion of these jobs are failed or terminated right after submission, thus they do not cause significant overhead to the computing resources.
In the end, we successfully extracted 627K (out of 670K) job samples from the first 28 days' trace data. 

\subsubsection{Backblaze disk stats dataset}

The Backblaze dataset contains the statistics of the hard drives in the Backblaze data center~\cite{backblaze2020drive}.
The dataset contains daily snapshots of operational hard drives in the data center, including drive information (e.g., model, disk capacity) and SMART (Self-Monitoring, Analysis, and Reporting Technology) statistics, where SMART is a manufacturer-implemented system for the monitoring and early detection of errors. 
Figure~\ref{fig:disk_schema} shows the dataset schema and information provided in the Backblaze disk stats dataset.
The Backblaze disk stats dataset contains hard drive monitoring data collected from 2013 to 2020. 
Initially, from 2013 to 2014, the trace captured 40 different SMART attributes; 
then, from 2015 to 2017, the number of SMART attributes increased to 45; 
starting from the fourth quarter of 2018, 62 different SMART attributes are in the data.
Despite the change of monitored attributes, the data type and the monitoring interval (i.e., daily) are kept consistent from 2013 to 2020.
We focus on the data collected from 2015 to 2017 because: 1) the subset contains a large number of samples (i.e., over 40M samples), and 2) the subset contains a fixed set of SMART attributes while the data in other periods contains less (before 2015) or more (after 2017) SMART attributes.

Following prior works~\cite{mahdisoltani2017proactive,botezatu2016predicting}, our goal on the Backblaze dataset is to predict hard drive failures (i.e., sector error) within a given future time period (i.e., one week) based on the monitoring data captured during a period of time (i.e., one week) in the past. 
We consider a disk fails if its ``sector error count'' SMART attribute increases (i.e., observe sector errors) in the given future time period, same as described in the prior work~\cite{mahdisoltani2017proactive}.
The SMART attributes in the Backblaze dataset can be categorized into two types: cumulative attributes whose values are accumulated counts over the disk's lifetime, such as the ``reallocated sectors count''; and noncumulative attributes whose values reflect only the current status, such as the ``read error rate''. 
Knowing the recent changes in cumulative attributes rather than their raw values might be more insightful.
Therefore, we capture both the value change in the past time period and the raw value in the last day of the one-week past window as features for cumulative attributes while only capturing the last day's value for noncumulative attributes.
As a result, we collect 11 features from the raw values and 8 features derived from the raw values' differences.
%The detailed list of our used features for disk failure prediction is described in Table~\ref{tab:features_backblaze}.
The collected features are the same as those used in prior work~\cite{mahdisoltani2017proactive}, the most predictive ones selected from all the traced SMART attributes; all 19 features are temporal features.
We then collect data samples along a sliding one-week time window and only track the disks that are alive during the whole time window.
As a result, we extract 41M samples from the daily snapshots between 2015 and 2017.

\subsubsection{Alibaba GPU cluster trace dataset}

The GPU cluster trace dataset from Alibaba provides traces of workloads collected from the operation of a large-scale data center~\cite{alibaba2020cluster}.
The trace data is collected from runtime information on over 6,500 GPUs across about 1,800 machines in a period of 2 months spanning from July to August of 2020~\cite{weng2022mlaas}.
The dataset features ML jobs submitted by various users. 
Once a user submits a job, the job is translated into multiple tasks of different roles. 
Subsequently, each task is then allocated as instances running on machines. 
Figure~\ref{fig:alibaba_schema} illustrates the trace schema and available information provided in the Alibaba trace dataset.
Similar to the Google cluster trace dataset, sensitive fields such as username and job name are desensitized to protect users' privacy.

We define the task as predicting job outcomes using the configuration information and performance metrics available in the first five minutes since job submission~\cite{lyu2024update}, similar to our case study on the Google cluster trace dataset. 
The dataset contains cluster monitoring data for a total of 69 days (around nine weeks). 
To avoid abnormality (e.g., truncated and untracked jobs) on data samples close to the beginning and end of the trace data, we initiate feature extraction from the fourth day since the trace starts and collect features for a total of 8 weeks.
Similar to our handling method on the Google cluster trace dataset, we also removed unfinished jobs and jobs ending in less than five minutes and extracted 701K out of the total of 1.26M jobs.

\subsubsection{Dataset segmentation}

%\yingzhe{Maybe merge this subsection with the following model training subsection? but I cannot think of a good section name for it.}\heng{I think the current structure is fine}
In this case study, we simulate the real-world scenario where AIOps models are trained on a chunk of initially available data samples and then deployed in the field to predict future samples coming in batches.
We follow the approach used in prior studies for updating AIOps models~\cite{lin2018predicting,li2020predicting,xu2018improving,lyu2024update} to partition the studied operation datasets. 
Specifically, we partition each studied dataset into multiple time periods based on the natural time intervals, as described below:

\begin{itemize}
    \item \textbf{Google cluster trace dataset.} We partition the entire 28-day trace data into 28 one-day time periods.
    \item \textbf{Backblaze disk stats dataset.} We partition the entire 3-year monitoring data into 36 one-month time periods.
    \item \textbf{Alibaba GPU cluster trace dataset.} We partition the 2-month GPU trace data into 8 one-week time periods.
\end{itemize}

For each dataset, we use samples from the first half of the time periods as the initially available training data and the second half of the time periods as the testing data, consecutively coming in by temporal order.

\subsubsection{Model training}

To ensure that the result of our case study is generalizable, we include a variety of machine learning classifiers that have been used in prior works for predicting disk failures on the Backblaze disk stats dataset and job failures on the Google and Alibaba cluster trace datasets~\cite{elsayed2017learning,mahdisoltani2017proactive,botezatu2016predicting,lyu2021empirical,lyu2024update} in our case study.
The list of models we use includes Logistic Regression (LR), Classification and Regression Trees (CART), Random Forest (RF), and Multi-Layer Perceptron Neural Network (NN).
% We skip the GBDT model that was used in our prior works as its performance and mechanisms are both similar to the RF model
We follow the same model configuration as recorded in prior work~\cite{rosa2015catching,mahdisoltani2017proactive,lyu2021empirical,lyu2024update}.
We train the models using a sliding window training set with the length fixed to half of the total time periods.
Whenever a new testing period concludes and before the testing on the next time period, we train a new sliding window model to maintain model performance against concept drift (i.e., periodical retraining).
%\yingzhe{Here, this part feel like related to both model training and data partition, and it is the reason I propose to merge the two subsections.}

The studied operation datasets can be extremely imbalanced, with only 1\% job failure in the Google dataset and 0.1\% hard drives failures in the Backblaze dataset. 
To mitigate the impact of imbalanced dataset on model performance, we downsample the majority class (i.e., succeed jobs in the Google dataset and normal hard drives in the Backblaze dataset) in the training dataset to a success-to-fail ratio of 10:1 prior to training the model.
For the Alibaba GPU cluster datasets, we do not applied downsampling to the training samples as the job failure rate is 34.5\%.
It is worth noting that although we rebalance the training dataset by downsampling the majority class, we do not perform such downsampling on testing dataset.
To mitigate the impact of varying feature scales in different time periods on model performance and interpretation, we perform data standardization on each metric in the training dataset by removing the median of the metrics and scaling the metrics according to the quantile range.
We apply the same data scaler that was fitted on the training dataset for the respective historical model when predicting testing samples. 

Prior works~\cite{tantithamthavorn2018parameter,tantithamthavorn2016automated,song2013impact} also suggest that hyperparameter settings can significantly impact the performance of prediction models.
Therefore, we tune the hyperparameters of our studied models on the training data using a random search. 
We choose a random search instead of a grid search for our hyperparameter tuning as using random search is more efficient and can find models that are as good as or better than using a grid search~\cite{james2012random}.

\subsection{Model selection mechanisms}

% \subsubsection{Studied model selection mechanisms}

%\yingzhe{We now in new RQ1 discuss the theoratical optimal performance of model selection. Then in RQ2 (the current RQ1), use the optimal, and the two baselines (with periodical retraining and stationary as two special ``selection''). In RQ3 instead of stability, we measure the ranking similarity to the optimal. As we measure the ranking, we would always choose the best one and discard the ``ensemble-like'' best three/five etc.(in reality, we did not discard the discussion, but include it in the Jaccard similarity with different cutoff points)}
%Prior work also suggest that naive averaging or majority voting that followed by most of today's multi-model decision-maker may not be optimal as model may be sensitive to different types of concept drift~\cite{zhang2025dynamic}. <- not relavent anymore
Due to the diverse underlying mechanisms that cause concept drift, the best methods for enhancing models' robustness against it differ across different datasets and shifts~\cite{zhang2025dynamic}.
Therefore, we assess the performance of various model selection mechanisms on recycling the previously trained historical models, including two labeling-free model selection mechanisms proposed in prior works~\cite{liu2022consistent,hu2024laf}, two model selection mechanisms leveraging temporal adjacency and feature similarity proposed by ourselves, and two variants on the aforementioned mechanisms utilizing the temporal adjacency or similarity.
When predicting samples from a target testing period, the model selection mechanisms first estimate the performance of candidate historical models on the testing samples and then rank the models based on the estimation.
%\heng{would be better to use ``target'' instead of ``future'', since when we apply models on the data, it becomes current, not future}

We define the problem of model selection on historical models as follows.
Given the set of $n$ historical models $S_\mathcal{M}=\{\mathcal{M}_1, \mathcal{M}_2, \dots, \mathcal{M}_n\}$ and the testing context $\mathcal{C}_t=\{s_1,s_2,\dots,s_m\}$ on time period $t$, where samples in $\mathcal{C}_t$ are unlabeled\footnote{For example, for the job failure prediction task, the samples in $\mathcal{C}_t$ contain information about the characteristics of the jobs (e.g., configurations, early running status), and we need to predict the job outcomes which are unknown (i.e., unlabeled).}, the task is to estimate the rank $r$ of the historical models in $S_\mathcal{M}$ according to their expected performance on the whole testing context $\mathcal{C}_t$ with as small rank error as possible.
Historical context $\mathcal{C}_1, \mathcal{C}_2, \dots, \mathcal{C}_{t-1}$ and their respective labels $\mathcal{L}_1, \mathcal{L}_2, \dots, \mathcal{L}_{t-1}$ are also available for the model selection mechanisms.

In our case study, we empirically evaluate the following model selection mechanisms on historical models. 
%\heng{probably need illustrative figures (like the ones used in the AIOps maintenance paper) for readers to understand the mechanisms}\yingzhe{I tried but it would be hard to illustrate these different in the mechanism}

\subsubsection{Temporal adjacency based selection mechanism (TBM)} 
Prior works utilize the temporal adjacency in training and estimation of model performance in the environment of rapidly changing data distribution~\cite{yao2019revisiting}.
We consider the samples from the last period of time to be the most similar to the testing samples in terms of data distribution and estimate the performance of historical models with their performance on the last available period.
We also add a revised version (rTBM) that further utilize the temporal adjacency: when the highest ranked model is from the second latest historical time periods, we assume the model from the latest historical time period (which was not tested due to the concern of data leakage) would be better and bring it to the top of the ranking.

\subsubsection{Similarity based selection mechanism (SBM)} 
Aside from utilizing temporal adjacency to approximate the most similar historical samples to the testing samples, we also devise a model selection mechanism to find the most similar samples from historical time periods to the testing samples directly using distance measurements.
Hausdorff distance~\cite{birsan2006one} measures how far two subsets of a metric space are from each other and represents the greatest of all the distances from a point in one set to the closest point in the other set.
We calculate the Hausdorff distance between the samples in the testing period and the samples from each of the historical periods starting from the ending period of the first training window and find the most similar historical period to validate the historical models.
We then rank the historical models based on the performance of the chosen time period (i.e., the most similar historical period to the testing period).
We further place the ranking of the historical models without data leakage (i.e., no intersection between the training window and the chosen time period) above the models with data leakage to mitigate the issue.
We choose to use the Hausdorff distance as other distance measurements cannot be applied to multivariate distributions with different amounts of observations (e.g., Hellinger distance) or are computationally intensive (e.g., Wasserstein metric).
We also provide a revised version (rSBM) that ignores the data leakage issue and ranks the historical models altogether.
    
\subsubsection{CRC model selection mechanism (CRC)} 
CRC is a labeling-free selection mechanism that uses the predicted probabilities on the testing samples to estimate model performance~\cite{liu2022consistent}.
It assumes that a candidate model with higher confidence should yield more accurate prediction results and better performance.
The CRC mechanism first obtains the predicted probabilities of testing samples from each historical model.
It then ranks the models based on the average confidence calculated from the highest predicted probabilities among all the classes on each sample from the testing period.

\subsubsection{LaF model selection mechanism (LaF)}
LaF is also a labeling-free model selection mechanism that ranks the candidate models by maximizing the expectations of each model on consensus-decided pseudo-labels of the testing samples~\cite{hu2024laf}.
The LaF mechanism first uses majority voting on the prediction results from candidate models to estimate the pseudo-label of the testing samples.
It then removes the testing samples in which all candidate models agree on the prediction results (i.e., where a unanimous agreement occurs) as these samples possess no discriminability against the candidates.
LaF then employs the expectation-maximization algorithm on the trimmed-down testing set to optimize the estimation of candidate model capabilities.
Finally, it uses the final estimation to rank the candidate models.
%\heng{``exclude'' or ``only keep''? ``exclude ... agree on...'' does not match ``consensus-decided'' mentioned above}

\subsubsection{Conventional baselines}
We consider two conventional baselines in our case study: stationary model and periodical retraining.
The two baselines are conventional model updating strategies that are widely used in prior AIOps 
works~\cite{elsayed2017learning,rosa2015catching,botezatu2016predicting,mahdisoltani2017proactive,chen2019outage,xue2018spatial,li2020predicting,dang2019aiops,lin2018predicting}.
A stationary model is trained on the initial training data and never updates while a periodical retraining mechanism updates the model with sliding window training samples each time data from a new period becomes available. 
The stationary model can be seen as a model selection mechanism that always selects the initially trained model while the periodical retraining can be seen as a model selection mechanism that always selects the newest model.
%Along with the multi-model selection mechanisms above, we also have two baselines: %The first baseline SM indicates a lower bound of the model performance while the second baseline SW indicates an optimal performance when constantly updating AIOps solutions in the field.

\subsubsection{Hypothetical oracle}
We also design a hypothetical oracle that exploits the true labels of samples from the target time period to rank and select the historical models.
During the inference of target samples, their label are not available yet hence the oracle shows a theoretical performance only.
%\heng{might be better to use ``prediction'' or ``inference'' instead of ``testing''; otherwise, need to explain why the labels are not available} samples
The hypothetical oracle first estimates the performance of candidate historical models with testing samples and the yet-to-come labels then ranks the candidates with the prediction performance estimation.
The oracle serves as a \emph{theoretical optimal} performance indicator for historical model selection mechanisms.
%\heng{it seems that using AUC as the evaluation metric is not defined yet? Only this oracle uses AUC or other selection mechanisms also use AUC? It reads strange to have AUC here only}

\subsection{Evaluation of model selection mechanisms}
We measure the performance of model selection mechanisms from three aspects: prediction performance, ranking performance, and ranking consistency.
For the prediction performance, we measure how well the top model chosen by each model selection mechanism can predict the incidents on testing samples.
For the ranking performance, we measure how similar the rankings from model selection mechanisms are to the actual model ranking by their performance on the testing samples.
We also measure the consistency of model rankings from the same model selection mechanism in different runs (i.e., with different random seeds).

\subsubsection{Prediction performance}
We evaluate the performance of our models using the Area Under the Receiver Operating Characteristic Curve (AUC) metric, which is a standard and widely used metric for evaluating machine learning models. 
AUC measures model performance by calculating the area under the curve of true positive rate (TPR) against false positive rate (FPR) at different classification thresholds.
Prior work~\cite{tantithamthavorn2018experience} shows that one should use threshold-independent metrics such as AUC in lieu of threshold-dependent metrics such as Precision, Recall, or F-measure to evaluate model performance. 
Therefore, in this case study, we use AUC as the performance indicator of prediction results on the testing samples.
%Because of the temporal nature of AIOps datasets, traditional evaluation methods such as cross-validation may result in data leakage and therefore lead to inaccurate performance evaluation~\cite{LyuTOSEM21}. 
%In this case study, we evaluate a model that is trained on a specific period of dataset using data in the next period as the testing dataset. 
We measure the AUC performance of each model selection mechanism on each testing period (i.e., from period $l/2+1$ to period $l$ for the $l$ time periods).
We repeat the experiment 100 times with different random seeds to mitigate the performance fluctuation.

To statistically compare the performance of different combinations of model update strategies and model choices, we apply the Scott-Knott test. The Scott-Knott test is a hierarchical clustering method~\cite{scott1974cluster} that groups observations into statistically distinct clusters. 
The observations within a group have no statistically significant difference (i.e., $p$-value $\geq$ 0.05), while the observations in different groups have a statistically significant difference (i.e., $p$-value $<$ 0.05).
In our case, the observations are the values of AUC from $100$ repeated experiments with different random seeds.

\subsubsection{Ranking Performance}
Model selection mechanisms rank the candidate models by their estimated performance on the testing samples.
To compare the similarity between the ranking of each model selection mechanism and the ground truth (i.e., ranking of models based on the true AUC performance on the testing samples), we use the following metrics:

% \hao{If I understand correctly, we are comparing the studied mechanisms with the oracle. For evaluation, we can look at top-1 agreement (measured by Kendall's $\tau$), top-n agreement (measured by Jaccard similarity), and robustness (measured by Kendall's W). Can we change the items below to `top-1 agreement', `top-n agreement', and `robustness'?}

% \begin{itemize}
\textbf{Kendall's $\tau$}. 
%\textbf{Overall agreement}. \yingzhe{$\leftarrow$ we better keep the names of metrics and explain the meanings in the text}
Kendall's $\tau$ measures the non-parametric rank correlation between two rankings for evaluating the overall agreement. 
Kendall's $\tau$ ranges from 0 (no agreement) to 1/-1 (perfect agreement/disagreement). 
% maybe not too mathematical
%Given $n$ historical models $S_\mathcal{M}=\{\mathcal{M}_1, \mathcal{M}_2, \dots, \mathcal{M}_n\}$, define ranking $r_i,j$ as the rank of model $i$ by mechanism $j$.
We focus on only the agreement of model rankings and use the same interpretation schema from prior work~\cite{rajbahadur2020impact} to interpret Kendall's $tau$ in our study: 
\[
\text{Kendall's}\ \tau\ \text{agreement} = 
\left\{ \begin{array}{ll}
  \text{Weak}      &  \text{if } 0 \leq \tau\leq 0.3. \\
  \text{Moderate}  &  \text{if } 0.3 < \tau \leq 0.6. \\
  \text{Strong}    &  \text{if } 0.6 < \tau \leq 1.
\end{array} \right.
\] 
In our case study, Kendall's $\tau$ measures the \emph{overall agreement} between rankings from model selection mechanisms and the ranking from the hypothetical oracle.

%\textbf{Top-k agreement} \yingzhe{$\leftarrow$ we better keep the names of metrics and explain the meanings in the text}
\textbf{Jaccard similarity coefficient $J_k$}. 
The Jaccard similarity coefficient $J_k$ evaluates the similarity between the top-$k$ model sets generated by the two rankings.
Prior work usually consider values of $k=3,5,10$~\cite{hu2024laf,liu2022consistent,meng2021measuring} to evaluate the ranking performance of model selection mechanisms on different cutoff points.
Given $n$ historical models $S_\mathcal{M}=\{\mathcal{M}_1, \dots, \mathcal{M}_n\}$ and two ranking functions $r_1$ and $r_2$ on models $S_\mathcal{M}$.
The Jaccard similarity coefficient on top-k models $J_k$ is defined as:
\[
J_k=\frac{|\cap_{i}\{\mathcal{M}_j|r_i(\mathcal{M}_j)\leq k\}|}{|\cup_{i}\{\mathcal{M}_j|r_i(\mathcal{M}_j)\leq k\}|}
\]
where $r_i(\mathcal{M}_j)\in \{1,...,m\}$, $i\in\{1,2\}$, and $j\in\{1,...,m\}$. 
A large $J_k$ indicates a higher degree of similarity between two rankings.
In our case study, the Jaccard similarity coefficient measures the \emph{agreement on top-k model candidates} between the rankings from model selection mechanisms and the ranking from the hypothetical oracle.
Since the Alibaba dataset only contains four testing periods (i.e., 4 historical models at maximum) while the prediction performance effectively measured the performance when $k=1$.
Therefore, in our case study, we only report the Jaccard similarity with $k=3$ (i.e., $J_3$).
We include the results for other cutoff points in our replication package.

%\yingzhe{Removed the robustness as the in-mechanism consistency is measured by the Kendall's W}

\subsubsection{Ranking consistency}
To measure the consistency of ranking results provided by each model selection mechanism on each testing period, we calculate Kendall's W.
Kendall's W is a non-parametric measure of the agreement among multiple rankings, ranging from 0 (no agreement) to 1 (complete agreement). 
In our case study, we calculate Kendall's W among the rankings from $100$ runs in each of the combinations of testing period, model, and dataset.
We use the same interpretation schema that is applied to Kendall's $\tau$ to evaluate the degree of agreement. 
%\heng{Not quite clear whether the consistency is for the same mechanism, different time periods, or the same period, different mechanisms. Would be good to give a clear/formal definition.}

We calculate these ranking performance metrics on each testing period starting from the second, as the first testing period contains only one historical model, which makes the rankings trivial. 
We also apply the Scott-Knott test to group the results from the 100 repeated experiments, providing a statistical basis for comparison.

% \textbf{Statistical ranking of the performance indicators} 
% For each of the above-mentioned evaluation metrics, we use the Scott-Knott test to rank the combinations of model update strategy and model choice statistically.
% Scott-Knott is a clustering technique~\cite{scott1974cluster} that groups observations into statistically distinct groups using hierarchical clustering analysis. 
% The observations within a group have no statistically significant difference (i.e., $p$-value $\geq$ 0.05), while the observations in different groups have a statistically significant difference (i.e., $p$-value $<$ 0.05).
% In our case, the observations are the evaluated metric values from $100$ repeated experiments with different random seeds.

\section{Experiment Results}
\label{sec:results}

We organize the experimental results along the evaluation of model selection mechanisms in the prediction performance of the top-ranked models, the ranking performance of model selection mechanisms when compared with the hypothetical oracle, and the ranking consistency inside each of the model selection mechanisms.
The three evaluation aspects correspond to the three RQs we aim to address.

%\heng{Would be better to link the subsection titles to RQs, otherwise it might be confusing to readers about what the answers to the RQs are}

%\subsection{Prediction performance}
\subsection{RQ1: How well can the model selection mechanisms achieve optimal performance for AIOps solutions?}

\begin{figure}[!ht]
    \centering
    \subfloat[Google]{\label{fig:auc_trend_google}{
        \includegraphics[width=0.95\textwidth]{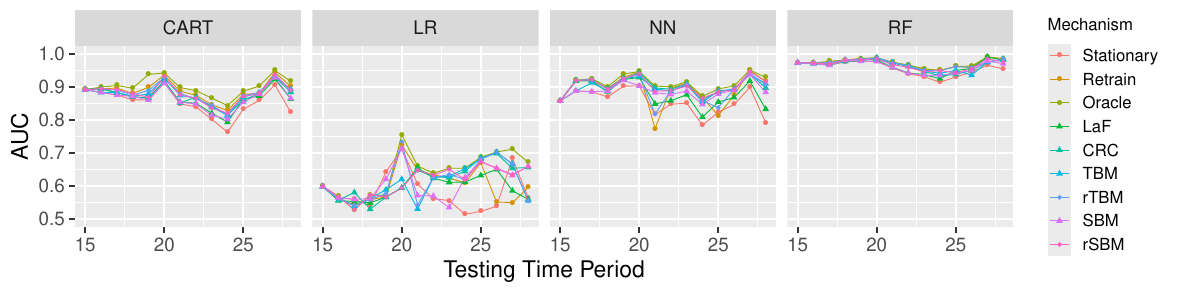}\hfill
    }}\hfill
    \subfloat[Backblaze]{\label{fig:auc_trend_backblaze}{
        \includegraphics[width=0.95\textwidth]{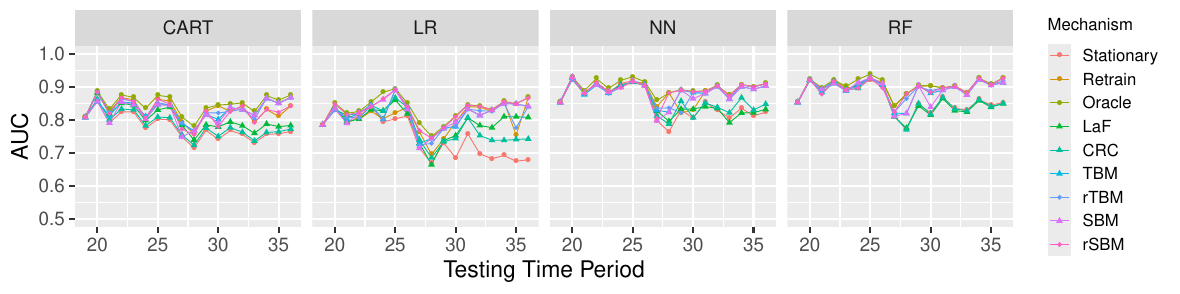}\hfill
    }}\hfill
    \subfloat[Alibaba]{\label{fig:auc_trend_alibaba}{
        \includegraphics[width=0.95\textwidth]{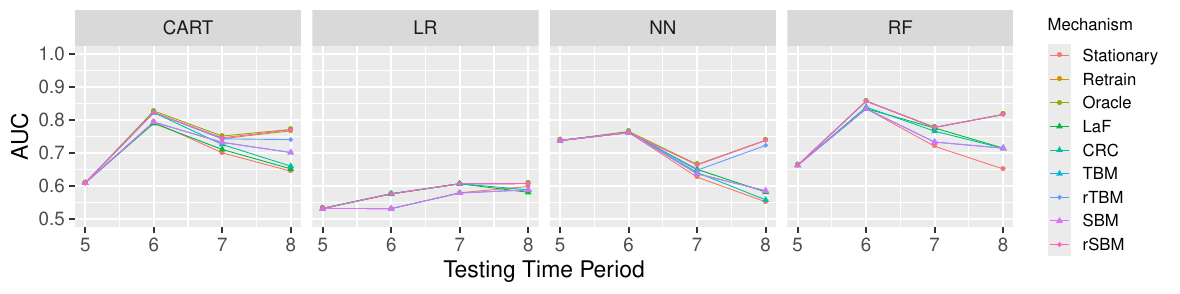}\hfill
    }}
    \caption{The average AUC performance of model selection mechanisms in each testing period.}
    \label{fig:auc_trend}
\end{figure}

\begin{figure}[!ht]
    \centering
    \subfloat[Google]{\label{fig:auc_sk_google}{
        \includegraphics[width=0.95\textwidth]{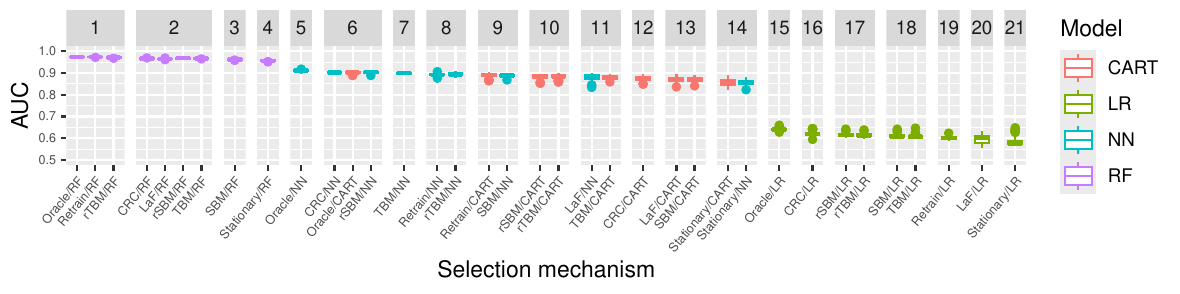}\hfill
    }}\hfill
    \subfloat[Backblaze]{\label{fig:auc_sk_backblaze}{
        \includegraphics[width=0.95\textwidth]{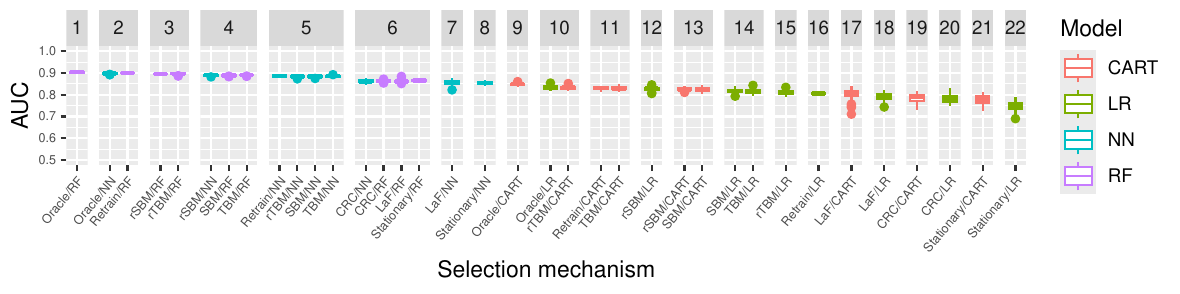}\hfill
    }}\hfill
    \subfloat[Alibaba]{\label{fig:auc_sk_alibaba}{
        \includegraphics[width=0.95\textwidth]{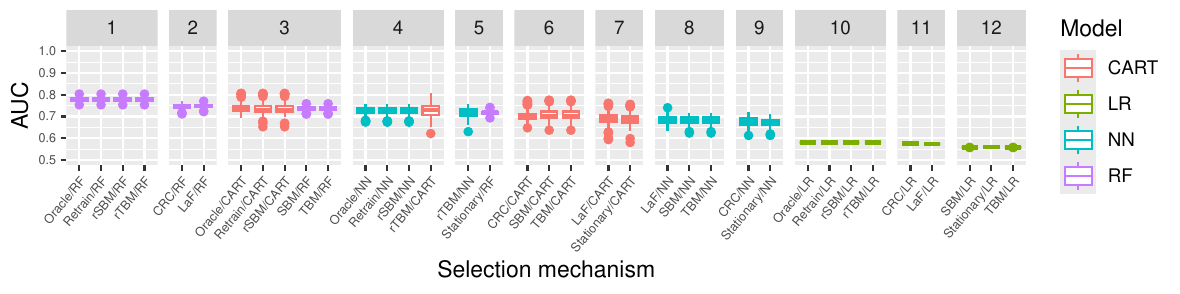}\hfill
    }}
    \caption{Scott-Knott test results of the AUC performance from different multi-model selection mechanisms.}
    \label{fig:auc_sk}
\end{figure}

Figure~\ref{fig:auc_trend} shows the AUC performance of each model selection mechanism using the top-ranked models in each testing period.
Figure~\ref{fig:auc_sk} further groups the AUC performance of each model selection mechanism using the Scott-Knott ranking test.

\textbf{The oracle shows higher performance than the periodical retraining baseline in many cases, indicating possible improvements over periodical retraining for model selection mechanisms.}
We observe higher AUC performance on the oracle baseline than the periodical retraining baseline in three models (i.e., NN, CART, and LR) from the Google dataset and all four models (i.e., RF, NN, CART, and LR) from the Backblaze datasets.
The observed performance differences are also statistically significant.
As the model selection mechanisms can choose from all historical models rather than being limited to the most recent one, these findings confirm the theoretical potential of model selection mechanisms in outperforming periodical retraining for maintaining AIOps solutions in the field.
However, we do not observe a significant performance difference between the oracle and periodical retraining on the Alibaba dataset.
This situation may result from the limited number of available historical models due to the relatively smaller volume of time periods and shorter trace data duration, which restricts concept drift from emerging.
% This situation may be related to the relatively small volume of time periods and short duration of the trace data, which restricts the amount of available historical models and hinders the concept drift from emerging.

%\textbf{All multi-model selection mechanisms except majority voting show higher performance than the stationary baseline}.
% <- this finding is pretty trivial?

\textbf{Several model selection mechanisms achieve higher performance than the periodical retraining baseline}.
% Google: CRC/NN SBM_Rev/NN TBM/NN | CRC/LR SBM_Rev/LR TBM_Rev/LR SBM/LR TBM/LR
% Backblaze: SBM_Rev/NN | TBM_Rev/CART | SBM_Rev/LR SBM/LR TBM/LR TBM_Rev/LR
% Alibaba: None
We observe that several model selection mechanisms achieve statistically better performance than the periodical retraining baseline on the Google and Backblaze datasets.
On the Google dataset, the CRC, rSBM, and TBM model selection mechanisms with the NN model, as well as the CRC, rSBM, rTBM, SBM, and TBM mechanisms with the LR model, achieve statistically better performance than the periodical retraining baseline.
On the Backblaze dataset, the rSBM mechanism with the NN model, the rTBM mechanism with the CART model, and the rSBM, SBM, rTBM, and TBM mechanisms with the LR model all achieve statistically better performance than the periodical retraining baseline.
However, we have not observed statistical performance differences on the Alibaba dataset due to the reasons mentioned above.

\textbf{The CRC, rSBM, and TBM mechanisms on the Google dataset, and the rSBM and rTBM mechanisms on the Backblaze dataset, show a generally better performance than other model selection mechanisms}.
On the Google dataset, the CRC, rSBM, and TBM mechanisms achieve better performance than the periodical retraining baseline with NN and LR models.
In addition, rTBM shows performance comparable to the periodical retraining baseline in RF, NN, and LR models.
On the Backblaze dataset, the rSBM and rTBM mechanisms achieve better performance than the periodical retraining baseline on two models each (NN and LR, CART and LR, respectively).
The SBM and TBM mechanisms also achieve comparable or higher performance to the periodical retraining on most models.
Although no model selection mechanism explicitly achieves superior performance compared to the periodical retraining baseline on the Alibaba dataset, rSBM and rTBM mechanisms present comparable results in most scenarios.
% On the Alibaba dataset, although no model selection mechanism achieve better performance than the periodical retraining baseline, the rSBM and rTBM mechanisms achieve comparable performance to the periodical retraining baseline on most of the models.

\textbf{The revised versions of the TBM and SBM mechanisms tend to have a better performance than the original ones}.
For the TBM mechanism, we observe that the revised version (i.e., rTBM) achieve better performance than the original one with three models (i.e., RF, CART, and LR) on the Google dataset, two models (i.e., RF and CART) on the Backblaze dataset, and all four models on the Alibaba dataset.
The revised version of TBM leverages the temporal adjacency to extend estimations to the most recent historical model, hence increasing the chance of selecting the best-performing model as the top-ranked choice.
For the SBM mechanism, the revised version (i.e., rSBM) achieves better performance than the original one with all four models on the Google dataset, three models (i.e., RF, NN, and LR) on the Backblaze dataset, and all four models on the Alibaba dataset.
We surmise that the rSBM mechanism has a greater chance of selecting more recent historical models when ignoring the intersection with training samples.
Such behavior further shows the weight of temporal adjacency when selecting historical models.

% compared with oracle: which achieved similar perf?
\textbf{The rTBM mechanism can achieve statistically comparable performance to the oracle baseline in some scenarios}.
We observe that the rTBM mechanism achieves comparable performance with the RF model on the Google dataset.
Although several model selection mechanisms also achieve similar performance to the oracle baseline on the Alibaba dataset, we do not consider these cases valid as the performance groupings are insufficiently discriminative due to the short duration.

\vspace{0.5cm}
\begin{Summary}{Summary of RQ1}{}
In terms of prediction performance using the top-ranked model, we observe several model selection mechanisms achieve higher performance than the periodical retraining baseline, with the rTBM mechanism achieving statistically comparable performance to the oracle baseline in some scenarios.
However, there is still a gap between the theoretical upper bound performance indicated by the hypothetical oracle and the performance of the existing model selection mechanisms.
\end{Summary}

\begin{figure}[!t]
    \centering
    \subfloat[Google]{\label{fig:kendal_tau_trend_google}{
        \includegraphics[width=0.95\textwidth]{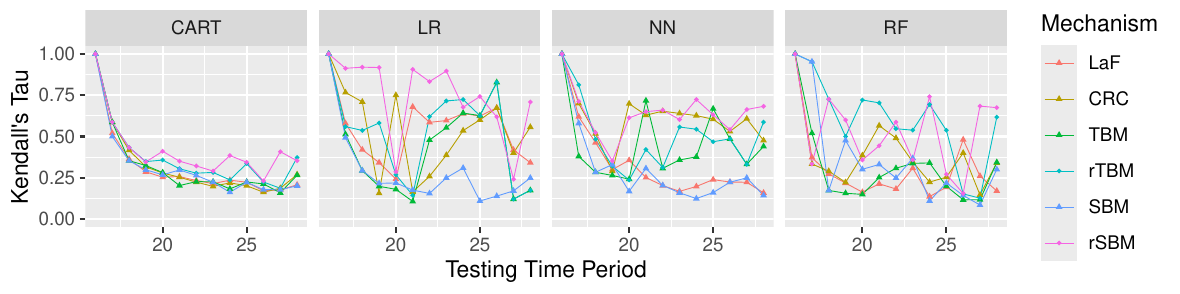}\hfill
    }}\hfill
    \subfloat[Backblaze]{\label{fig:kendal_tau_trend_backblaze}{
        \includegraphics[width=0.95\textwidth]{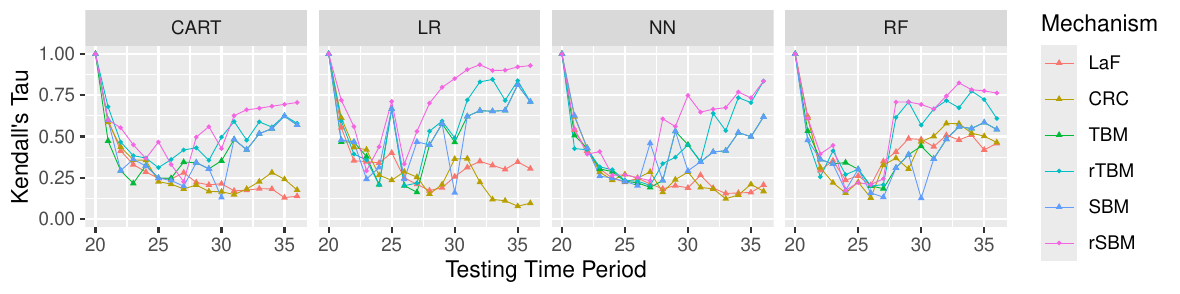}\hfill
    }}\hfill
    \subfloat[Alibaba]{\label{fig:kendal_tau_trend_alibaba}{
        \includegraphics[width=0.95\textwidth]{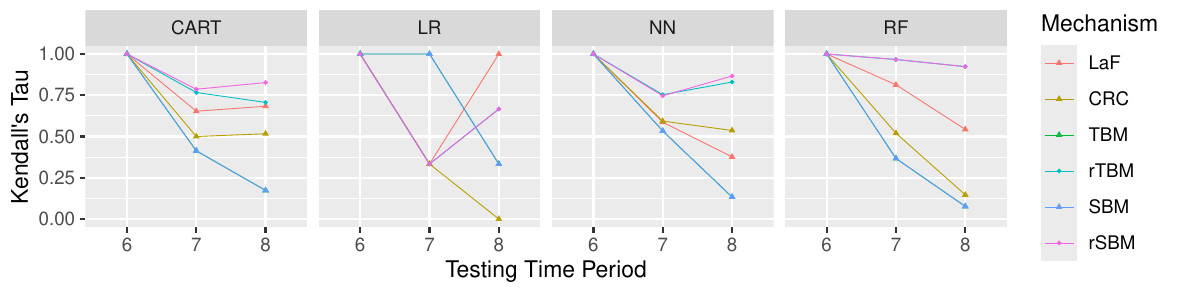}\hfill
    }}
    \caption{The average Kendall's $\tau$ correlation between model selection mechanisms and the oracle ranking in each testing period.}
    \label{fig:kendal_tau_trend}
\end{figure}

\begin{figure}[!t]
    \centering
    \subfloat[Google]{\label{fig:kendal_tau_sk_google}{
        \includegraphics[width=0.95\textwidth]{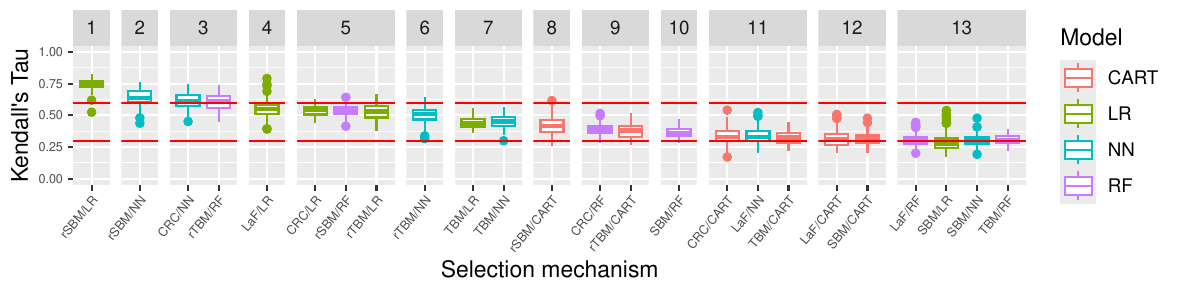}\hfill
    }}\hfill
    \subfloat[Backblaze]{\label{fig:kendal_tau_sk_backblaze}{
        \includegraphics[width=0.95\textwidth]{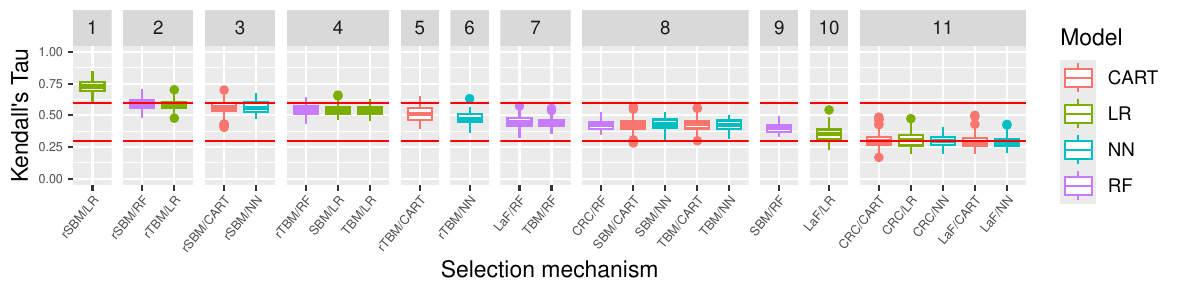}\hfill
    }}\hfill
    \subfloat[Alibaba]{\label{fig:kendal_tau_sk_alibaba}{
        \includegraphics[width=0.95\textwidth]{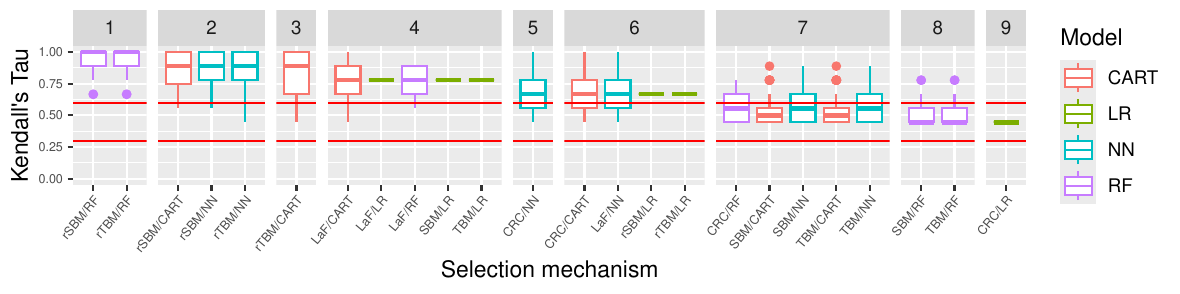}\hfill
    }}
    \caption{Scott-Knott test results of the Kendall's $\tau$ correlation from different multi-model selection mechanisms. The horizontal lines indicate the threshold for interpreting ranking agreement.}
    \label{fig:kendal_tau_sk}
\end{figure}

\subsection{RQ2: How well can the model selection mechanisms achieve optimal model ranking for AIOps solutions?}
%\subsection{Ranking Performance}

\begin{figure}[!t]
    \centering
    \subfloat[Google]{\label{fig:jaccard_trend_google}{
        \includegraphics[width=0.95\textwidth]{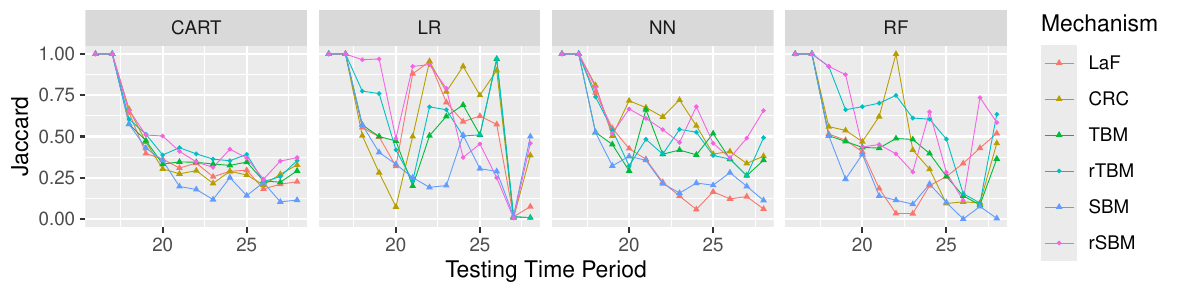}\hfill
    }}\hfill
    \subfloat[Backblaze]{\label{fig:jaccard_trend_backblaze}{
        \includegraphics[width=0.95\textwidth]{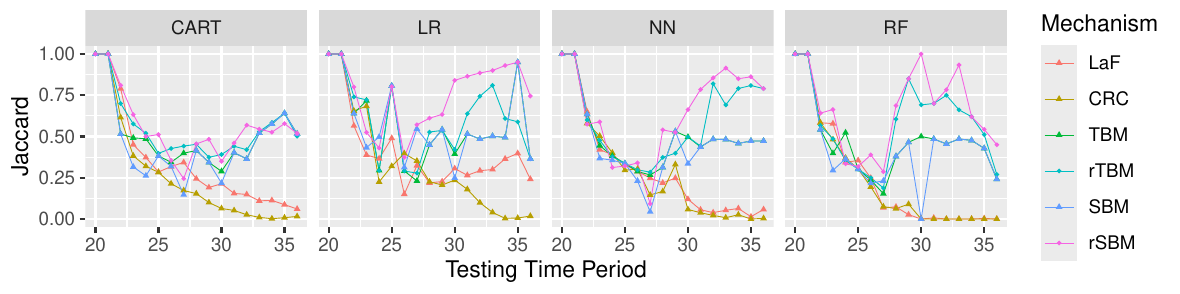}\hfill
    }}
    % \hfill
    % \subfloat[Alibaba]{\label{fig:jaccard_trend_alibaba}{
    %     \includegraphics[width=0.95\textwidth]{figures/selection_model_jaccard_trend_alibaba.pdf}\hfill
    % }}
    \caption{The Jaccard similarity coefficient ($k=3$) between model selection mechanisms and the oracle ranking in each testing period.}
    \label{fig:jaccard_trend}
\end{figure}

\begin{figure}[!t]
    \centering
    \subfloat[Google]{\label{fig:jaccard_sk_google}{
        \includegraphics[width=0.95\textwidth]{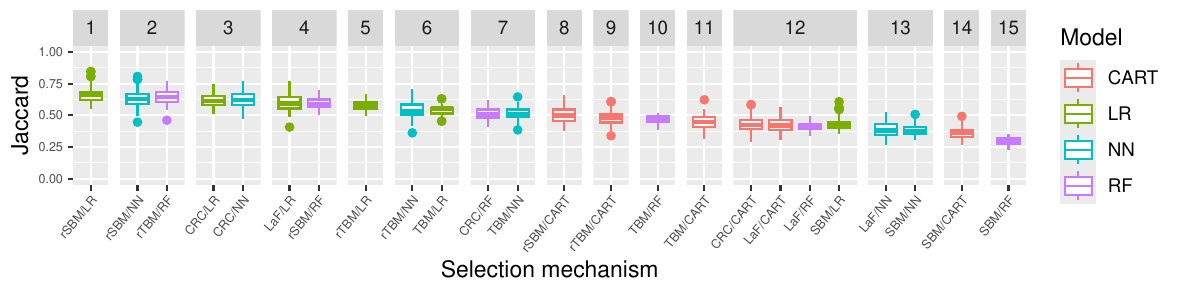}\hfill
    }}\hfill
    \subfloat[Backblaze]{\label{fig:jaccard_sk_backblaze}{
        \includegraphics[width=0.95\textwidth]{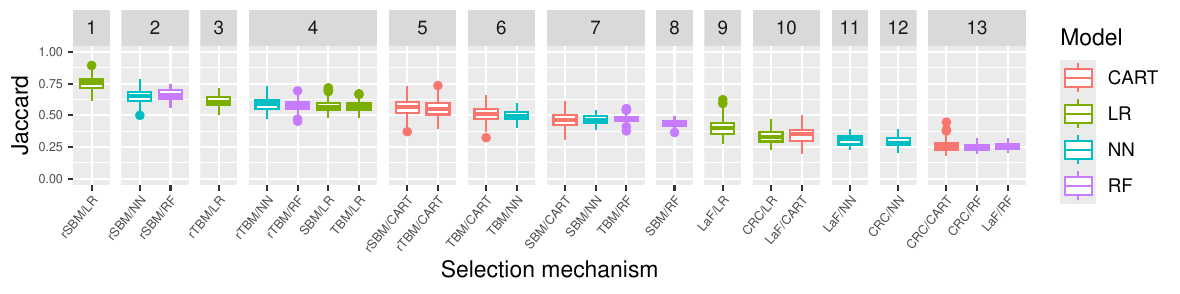}\hfill
    }}
    % \hfill
    % \subfloat[Alibaba]{\label{fig:jaccard_sk_alibaba}{
    %     \includegraphics[width=0.95\textwidth]{figures/selection_model_jaccard_sk_alibaba.pdf}\hfill
    % }}
    \caption{Scott-Knott test results of the Jaccard similarity coefficient ($k=3$) from different multi-model selection mechanisms.}
    \label{fig:jaccard_sk}
\end{figure}

We measure the similarity between the rankings from model selection mechanisms and the oracle baseline using Kendall's $\tau$ and Jaccard similarity coefficient as indicators of ranking performance. We also calculate Kendall's W among the rankings from the same model selection mechanism in each testing period to measure the consistency of rankings.
For Kendall's $\tau$, Figure~\ref{fig:kendal_tau_trend} shows the Kendall's $\tau$ between the rankings on each model selection mechanism and the oracle baseline in each testing period and Figure~\ref{fig:kendal_tau_sk} further groups the performance of each model selection mechanism using the Scott-Knott ranking test.
Similarly, for the Jaccard similarity coefficient, Figure~\ref{fig:jaccard_trend} shows the Jaccard similarity coefficient between the rankings on each model selection mechanism and the oracle baseline in each testing period and Figure~\ref{fig:jaccard_sk} further groups the performance of each model selection mechanism using the Scott-Knott ranking test.

\textbf{The rSBM model selection mechanism provides rankings most aligned with the oracle baseline across datasets and models}.
Figure~\ref{fig:kendal_tau_trend} and Figure~\ref{fig:kendal_tau_sk} show that the rSBM model selection mechanism consistently achieves the highest Kendall's $\tau$, which indicates strong ranking similarity to the oracle. Specifically, rSBM is ranked within the top statistical group for three (i.e., LR, NN, and CART) of the four models on the Google dataset. Also, rSBM achieves the highest statistical group ranking for all four models on the Backblaze dataset, and for three (i.e., RF, CART, and NN) models on the Alibaba dataset.
The rTBM model selection mechanism similarly ranks among the top group in terms of Kendall's $\tau$. For example, rTBM achieves the top group for the RF model on the Google dataset, for the LR model on the Backblaze dataset, and for the RF and NN models on the Alibaba dataset.

\textbf{The rSBM and rTBM model selection mechanisms perform well in terms of top-k agreement.}
As shown in Figure~\ref{fig:jaccard_trend} and Figure~\ref{fig:jaccard_sk}, we observe similar performance for the rSBM and rTBM model selection mechanisms based on Jaccard similarity coefficient with $k=3$.
On the Google dataset,  the rSBM model selection mechanism is within the top statistical performance group for LR and CART, while rTBM achieves the highest statistical group ranking for the RF model. For the Backblaze dataset, rSBM maintains top-level performance across all four models, while rTBM achieves the top statistical group ranking for the LR and CART models.
We omit the Alibaba dataset from this analysis due to the limited number~(i.e., four) of testing periods, which makes Jaccard similarity analysis impractical (i.e., fewer historical models than required by the selected top-k threshold, resulting in trivial or total agreement for early periods).
% ~\yingzhe{Should we remove the Jaccard figures for the Alibaba dataset?} \hao{Yes, I think so. However, I’m wondering what key takeaway we should emphasize regarding the results of the Jaccard similarity. Specifically, what additional insight does it provide compared to reporting only Kendall's $\tau$.}~\yingzhe{The Jaccard similarity strengthens the verdicts on correlation, prior works usually report both.}

\begin{figure}[!ht]
    \centering
    \subfloat[Google]{\label{fig:kendal_w_trend_google}{
        \includegraphics[width=0.95\textwidth]{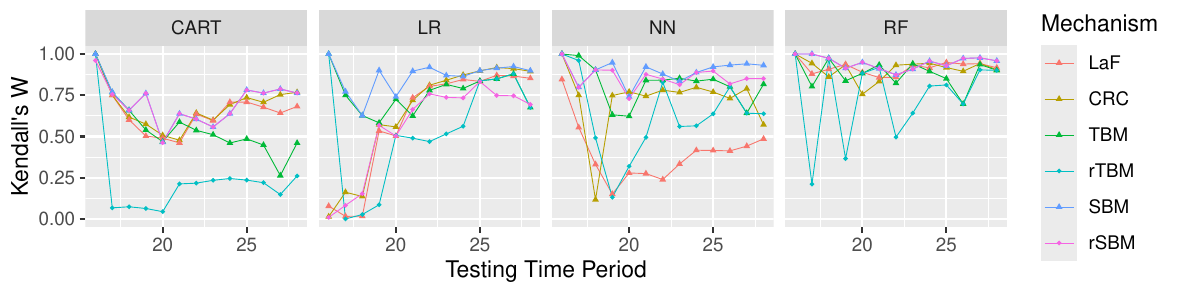}\hfill
    }}\hfill
    \subfloat[Backblaze]{\label{fig:kendal_w_trend_backblaze}{
        \includegraphics[width=0.95\textwidth]{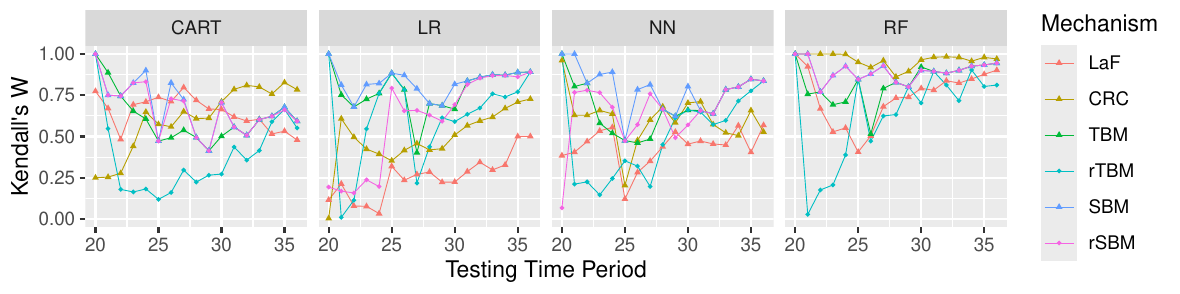}\hfill
    }}
    % \hfill
    % \subfloat[Alibaba]{\label{fig:kendal_w_trend_alibaba}{
    %     \includegraphics[width=0.95\textwidth]{figures/selection_model_w_trend_alibaba.pdf}\hfill
    % }}
    \caption{The Kendall's W correlation among rankings from the same model selection mechanism in each testing period.}
    \label{fig:kendal_w_trend}
\end{figure}

\begin{figure}[!ht]
    \centering
    \subfloat[Google]{\label{fig:kendal_w_sk_google}{
        \includegraphics[width=0.95\textwidth]{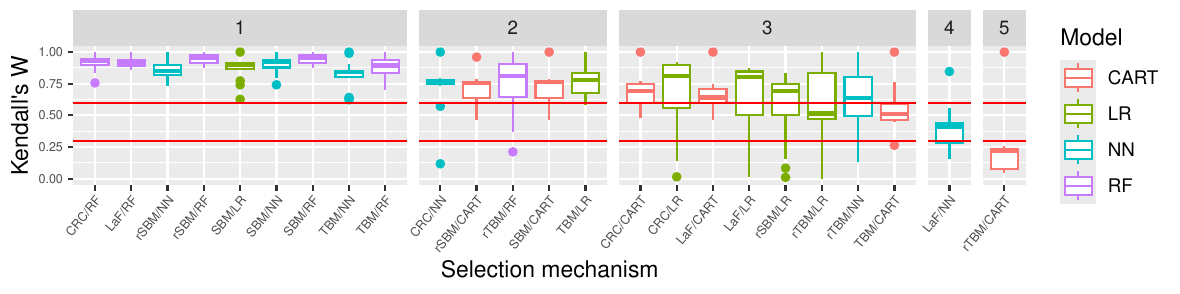}\hfill
    }}\hfill
    \subfloat[Backblaze]{\label{fig:kendal_w_sk_backblaze}{
        \includegraphics[width=0.95\textwidth]{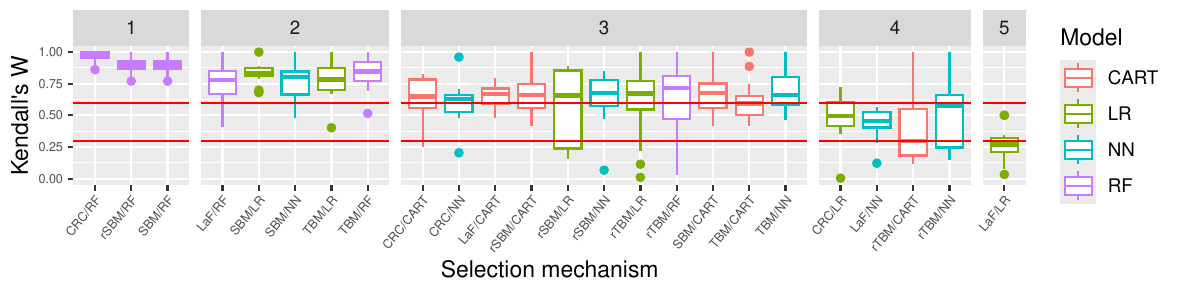}\hfill
    }}
    % \hfill
    % \subfloat[Alibaba]{\label{fig:kendal_w_sk_alibaba}{
    %     \includegraphics[width=0.95\textwidth]{figures/selection_model_w_sk_alibaba.pdf}\hfill
    % }}
    \caption{Scott-Knott test results of the Kendall's W correlation from different multi-model selection mechanisms. The horizontal lines indicate the threshold for interpreting ranking agreement.}
    \label{fig:kendal_w_sk}
\end{figure}

\vspace{0.5cm}
\begin{Summary}{Summary of RQ2}{}
We measure the similarity between the rankings from model selection mechanisms and the oracle baseline using Kendall's $\tau$ and Jaccard similarity coefficient $J_3$ as indicators of ranking performance.
Our results find that the rSBM model selection mechanism provides rankings most aligned with the oracle baseline across datasets and models, while the rSBM and rTBM model selection mechanisms perform well in terms of the agreement on the top-3 performing models.
\end{Summary}

%\heng{robustness/consistency is better to be in a separate sub-section, to correspond to the RQs}
\subsection{RQ3: How stable are the model ranking results achieved by the model selection mechanisms?}
Figure~\ref{fig:kendal_w_trend} shows the Kendall's W among the rankings from each model selection mechanism in each testing period, and Figure~\ref{fig:kendal_w_sk} further groups the Kendall's W of each model selection mechanism using the Scott-Knott ranking test.

\textbf{Overall, most model selection mechanisms achieve robust consistency with strong agreement among rankings across repeated runs}.
As shown in Figure~\ref{fig:kendal_w_trend} and Figure~\ref{fig:kendal_w_sk}, we observe strong agreement (i.e., Kendall's W $>$ 0.6) for nearly all combinations of model selection mechanisms and models on the Google dataset, with a few combinations demonstrating moderate or weak agreement. 
The rTBM model selection mechanism for the LR model, TBM for the CART model, and LaF for the NN model achieve moderate agreement, while rTBM for the CART model demonstrates weak agreement.
On the Backblaze dataset, the majority of combinations also show strong consistency, with four model and selection mechanism combinations showing moderate agreement, and one combination showing weak agreement (i.e., the LaF mechanism on the LR model). 
For the Alibaba dataset, we omit Kendall's W analysis due to the limited number of historical models and testing periods available.

\vspace{0.5cm}
\begin{Summary}{Summary of RQ3}{}
We evaluate the ranking consistency inside the model selection mechanism in each testing period.
In other words, we measure how the ranking provided by the same model selection mechanism with the same configuration varies from time to time to assess the trustworthiness of the rankings.
We observe that most model selection mechanisms achieve robust consistency with strong agreement among rankings across repeated runs, with the LaF and rTBM model selection mechanisms showing weak agreement among multiple runs in some cases.
\end{Summary}

% on the CART model and the LaF mechanism on the NN model achieve a moderate agreement while the rTBM mechanism achieve a weak agreement on the CART model on the Google dataset.
% On the Backblaze dataset, four model and selection mechanism combinations show moderate agreement, one combination shows weak agreement (i.e., LaF mechanism on the LR model), while all other combinations show strong agreement.
% Similar to the Jaccard similarity coefficient, we omit the results on the Alibaba dataset due to the small amount of historical models.

% \hao{Should we add a discussion about the potential to achieve oracle-level performance and include a call for future research?}

%\heng{Take-home message are not quite clear. Maybe two solutions: 1) add take-home messages at the end of each RQ sub-section, or 2) add an implication section to discuss the overall take-home messages: hypothetical oracle can achieve better performance than periodical retraining, suggesting that ...; XX methods are most effective at selecting ...; better performance does/does not suggest better robustness/consistency, etc. Each message with some supporting arguments/evidence.}

%\section{Discussion}
%\label{sec:discussion}
%\input{texts/05-Discussion.tex}

\section{Threats to Validity}
\label{sec:threats}
% \yingzhe{Hao can you help adding more content to this section? I feel like it is too short.}

\subsection{External Validity}
One possible threat to external validity is that we only examine the multi-model selection mechanisms on a limited number of datasets. 
To address this validity, we choose three large-scale and publicly available datasets that represent diverse real-world scenarios (Google cluster trace, Backblaze disk stats, and Alibaba GPU cluster trace). 
These three datasets have been widely used in prior AIOps research~\cite{elsayed2017learning,rosa2015predicting,botezatu2016predicting,xu2018improving,mahdisoltani2017proactive}.

Another potential threat arises from the choice of machine learning models. We employ four representative machine learning models (i.e., Logistic Regression, CART, Random Forest, and Neural Network) that have been used on the studied datasets from prior works~\cite{elsayed2017learning,mahdisoltani2017proactive,botezatu2016predicting,lyu2021empirical,lyu2024update}  to ensure fair comparison and generalizability of our results. However, it remains unclear whether our findings directly extend to other types of models or architectures that are not studied in this work.

In addition, the choice of model selection mechanisms could be a potential threat. To address this, we include six model selection mechanisms: two labeling-free mechanisms proposed in prior works~\cite{liu2022consistent,hu2024laf}, as well as our proposed TBM and SBM variants (and their revised versions) leveraging temporal adjacency and feature similarity. These selection mechanisms ensure our study covers diverse approaches.
% two model selection mechanisms and two corresponding revisions utilizing temporal and spatial adjacency proposed by ourselves in the case study.
%We choose to use RF as the base classifier as it has one of the best performance on the tested datasets~\cite{lyu2021empirical,lyu2024update}.

\subsection{Internal Validity}
One possible threat to internal validity concerns the measurement of performance indicators (i.e., AUC for prediction performance, Kendall's $\tau$ and Jaccard correlation for ranking performance, and Kendall's W for ranking consistency).
Considering various factors may impact the model performance, we repeat the experiment for $100$ runs and control the randomness during each run.
The choice of time period partition may also pose a threat to internal validity.
We mitigate this threat by using the widely accepted natural time period partition as done in prior works~\cite{xu2018improving,li2020predicting}.

Furthermore, dataset preprocessing steps such as the downsampling strategy applied for handling class imbalance and data standardization may influence results. We adopt best practices widely used in prior empirical research~\cite{tantithamthavorn2018experience,elsayed2017learning,lyu2021empirical}. Specifically, we follow natural time intervals as recommended in prior AIOps research~\cite{xu2018improving,li2020predicting}, use standard approaches for handling data imbalance (downsampling the majority classes), and apply preprocessing steps such as data scaling according to median-quantile ranges.

\subsection{Construct Validity}
A threat to construct validity concerns our model training process, as the configuration and parameter settings may impact the experiment results. 
In order to mitigate this threat to construct validity, we use automated hyperparameter tuning to optimize the configuration of the hyperparameters, which is a widely used practice in the development of machine learning models. 

There may be other potential threats concerning our model training process. 
We use the same features as in prior works for the Google and Backblaze operation datasets and the same types of base models (except for the online learning approaches) have been applied in the prior works~\cite{rosa2015predicting,elsayed2017learning,botezatu2016predicting,mahdisoltani2017proactive} to reflect the training process in AIOps solutions.
Future work that evaluates our study by considering other modeling processes, e.g., using different features, different ML libraries, or a different re-sampling approach, could benefit our study.

\section{Conclusion}
\label{sec:conclusion}
In this work, we study the model selection mechanisms on historical models in the context of AIOps through a case study on three large-scale operation datasets and six types of model selection mechanisms.
We empirically evaluate the prediction performance when choosing the top-ranked model using model selection mechanisms, the ranking agreement between rankings from the model selection mechanisms and the theoretical oracle, and the ranking consistency inside each model selection mechanism in multiple runs.
Our findings suggest the temporal adjacency based selection mechanism tends to have a better performance than other model selection mechanisms and prevails in AUC performance than the periodical retraining.
We also find that the rSBM mechanism tend to have the most similar ranking when compared with the oracle ranking.
Future work may consider utilizing the ranking from rSBM mechanism further in scenarios like selecting models for time-based ensemble models considering its high accuracy.
We also suggest future research to devise model selection mechanisms that can achieve closer performance to the theoretical optimal as our results indicate there is still a gap in the performance between the current model selection mechanisms and the theoretical upper bound.

\section*{Declarations}
\label{sec:declarations}
\subsection*{Funding}
Not applicable.

\subsection*{Ethical Approval}
This study does not involve human participants or animals.

\subsection*{Informed Consent}
Not applicable. No human subjects were involved in this study.

\subsection*{Author Contributions}
\begin{itemize}
    \item Yingzhe Lyu: Conceptualization, Data Collection, Methodology, Data Analysis, Writing – Original
Draft.
    \item Hao Li: Methodology, Data Validation, Writing – Review \& Editing.
    \item Heng Li: Supervision, Writing – Review \& Editing, Conceptual Guidance, Research Direction.
    \item Ahmed E. Hassan: Supervision, Research Direction.
\end{itemize}

\subsection*{Data Availability Statement}
The dataset, experiment code, and experiment results of this study are available in our replication package.\footnote{
\url{https://github.com/EmpyreanKnight/suppmaterial-25-yingzhe-AIOpsSelection}}

\subsection*{Conflict of Interest}
The authors declared that they have no known competing interests or personal relationships that could have (appeared to) influenced the work reported in this article.
%All authors certify that they have no affiliations with or involvement in any organization or entity with any financial interest or non-financial interest in the subject matter or materials discussed in this manuscript.

\subsection*{Clinical Trial Number in the Manuscript}
Not applicable.

% BibTeX users please use one of
%\bibliographystyle{spbasic}      % basic style, author-year citations
\bibliographystyle{spmpsci}      % mathematics and physical sciences
\bibliography{aiops}

\end{document}